\documentclass[12pt,preprint]{aastex}

\slugcomment{Accepted: AJ, March 22, 2010}

\shorttitle{Sigma One}
\shortauthors{Madore}

\begin{document}


\title{Sigma One}


\author{Barry F. Madore}
\affil{Observatories of the  Carnegie Institution of Washington \\ 813 Santa 
Barbara St., Pasadena, CA ~~91101}
\email{barry@obs.carnegiescience.edu}



\begin{abstract}
We demonstrate that it is possible to calculate not only the mean of
an underlying population but also its dispersion, given only a single
observation and physically reasonable constraints (i.e., that the
quantities under consideration are non-negative and bounded). We
suggest that this counter-intuitive conclusion is in fact at the heart
of most modeling of astronomical data.

\end{abstract}

\vfill\eject

``Probability theory is nothing but common sense reduced to
calculation.''

\centerline{P.S. Laplace (1819) as quoted by E.T. Jaynes (2003)}

\section{INTRODUCTION}
In their discussion of optimizing search strategies for the
simultaneous discovery and characterization of variable stars using
predetermined but highly non-uniform sampling, Madore \& Freedman
(2005) investigated the small-sample limits. Their emphasis was on
situations where only a handful of observations (2-20 say) might be
available from which it would be necessary to determine periods,
measure mean magnitudes and derive full amplitudes. In this paper we
discuss how it is possible use a single observation to derive
confidence intervals on the mean (i.e., limits on variability, or
estimates of the population variance, or its full amplitude, etc.).

Shortly after the original manuscript was completed it was pointed out
that similar arguments have been made much earlier by Gott
(1993). Gott was interested in the estimation of ages and lifetimes of
events based single instances; and he presented confidence intervals
for the longevity of an object given a single observation of its
present age. He assumed a flat prior (of finite duration), and took
the stance that there cannot be anything special about any given time
that a random observation is made of an object that has a finite
lifetime. We return to a contextual discussion of this example, and a
generalization of it to modeling, at the end of this paper.

\section{ONE OBSERVATION}

What information can be derived about the parent population from a
single, isolated (non-negative) observation?  Here we are clearly
working at the observational information limit, and so a number of
plausible assumptions (``priors'') will of necessity enter the
calculations.  A tacit assumption is that the observation in question
is drawn from an underlying population having some
dispersion\footnote{Of course, if we assumed that the underlying
population had no dispersion our task would already be completed. To
first order most observers make this default assumption, modulo
observational error.} that we are trying to estimate.  We also assume
that this is a physical object and that all values (observed or
considered in the calculation) are therefore non-negative.

Lacking any information to the contrary we take the single observed
data point $y_1$ to be the best available estimate of the true mean, $<Y>$,
thus $<Y> = y_1$. It follows directly from the non-negativity
assumption that the maximum possible semi-amplitude of $Y$ is $A/2 = (<Y> -
0.0) = y_1$.  Invoking a symmetric prior would give a full amplitude
$A = 2 \times y_1$. Having given the flavor of this argument in the
above few lines we now more closely investigate a variety of priors,
but first we introduce a new tool, the {\it ``g-factor''}.

We ask the following question: if random samples, $y_i$, are drawn
from the parent distribution and considered one at a time, what is the
auxiliary distribution of multipliers, {\it ``g-factors''}, that would
convert the observation itself into its own ``error'', ($\epsilon_i$),
where $\epsilon_i$ is defined as being the absolute value of the
distance of the data point from the true population mean (i.e.,
$\epsilon_i \equiv \mid <Y> - y_i \mid$).  By this definition $g_i
\times y_i = \epsilon_i$ and so $g_i = \mid<Y>/y_i - 1.0\mid$. With
the {\it g-factor} distribution in hand, for any given prior
distribution, it is then possible to calculate specific multipliers
(g(50.0), g(68.5), g(90.0), {\it etc.}, by integrating over the {\it
g-factor} distribution function) that will guarantee that set
fractions (50.0\%, 68.5\%, 90.0\%, {\it etc.})  of the dataset will
fall within the range $y_1 \pm g(50.0) \times y_1, y_1 \pm g(68.5)
\times y_1,$ {\it etc.}, where $y_1$ is again our estimate for 
$<Y>$.

\section{PRIORS}

We now consider five realizations of four plausible underlying population
distribution functions -- a Poisson distribution, a uniform
distribution, an exponential, and two Gaussians.

\subsection{Poisson Distribution}

In this first example, invoking the Poisson distribution as being the
appropriate form for the underlying parent distribution, it is almost
trivially simple to demonstrate that a single observation can deliver
both an average and a meaningful value for the dispersion of the
underlying distribution. We first recall that the Poisson distribution
for rare events has the form $P(y) = \lambda^y e^{\lambda}/y!$ where
it is well known that the distribution is determined by a single
variable $\lambda$, and that $\lambda$ is numerically equal to both
the mean and the variance of this distribution. Therefore, if we have
a single observation $y_1$, which we assume comes from a Poisson process
then, as a first approximation we can equate $y_1$ with the mean value
for that population $<Y> \equiv \lambda = y_1 $, and by invoking the
known (definitional) equality of the mean and the variance for a
Poisson process, we have therefore also measured the
variance\footnote{And while not putting to fine a point to it, we also
note that this single observation also determines several higher
moments including the {\it skew} $= \lambda^{-1/2}$ and the {\it
kurtosis} = $\lambda^{-1}$}, with only a single observation.

\subsection{Uniform Parent Population}

We now consider a uniform distribution spanning the non-negative
interval [0,2] (Figure 1, upper panel).  The distribution of {\it
g-factors} is expected to be highly skewed: points near zero will
require very large factors to make their error bars overlap with the
true mean; however, all points greater than the mean (that is,
fully half of the ensemble for a uniform prior) will have calculated
error bars that overlap the mean for multiplicative {\it g-factors}
that are all less than 0.5.

We first solved for the {\it g-factor} distribution function by
simulation. We successively drew 100,000 observations from a uniform
distribution, calculated the factor that needed to be applied to that
number such that $y_1 \pm g_u\times{y_1}$ just overlaps the true mean
$<Y_u>$. In Figure 1 the results of that computer simulation are
shown, where the distribution function of multiplicative {\it g-factors}  is
found in the lower panel, and the parent distribution of individual
observations going into the simulation is given in the upper
panel. The lower panel also shows where the 50, 68.5, and 90\%
confidence intervals are found for the {\it g-factor} distribution
corresponding to this uniform prior.

If the underlying population, from which a single data point $y_1$ is
drawn, is itself uniformly populated and non-negative then $y_1 \pm
0.41\times{y_1}$ will contain the mean of the parent population 50\% of
the time. Other selected confidence intervals are given in Table 1.

After the computer simulations were completed it proved possible for
us to derive a closed-form analytic solution for multiplicative
factors, $g_u$ as a function of the confidence intervals ($CI$)
associated with this case of a uniform prior. We give these solutions
below; they are based on a simple mapping of the uniform distribution
into its {\it g-factors} (using the absolute values of the {\it
g-factors} accounts for the curious shape of the distribution in the
lower panel of Figure 1). The confidence intervals are then found by
integrating the normalized functional form of that mapping up to
required value of $CI$.

$$g_u = (\sqrt{1.0 + 4(CI)^2} - 1.0)/(2{CI}) \ \ \ \   [CI \le 2/3] $$

$$g_u = (1.0 - 2{CI})/(2(CI - 1.0))  \ \ \ \  [CI \ge 2/3] $$

\noindent
Exemplary values for common confidence intervals for the uniform prior
are given in the second column of Table 1. For the uniform prior they
were calculated from the analytic solution and are confirmed by the
simulations; all other soultions for other priors were derived from
the computer simulations.

\subsection{Exponential Parent Population}

Another plausible parent population is the exponential distribution.
Here we investigate the {\it g-factors} corresponding to that
distribution function. We have chosen to simulate an exponential
distribution with mean $<Y_e> = 1.0$. The results of the inversion are
shown in Figure 2 and the third column of Table 1. The contrast
between this and the uniform prior is clear. Because significantly
more samples will come with values close to zero their contribution to
the distribution of multiplicative factors will increase. This forces
the individual {\it g-factors} for specific confidence intervals to higher
values as compared to the uniform prior. Inspection of Table 1
confirms this quantitatively for all confidence intervals listed.

\subsection{Two Gaussian Distributions}

Since the uniform and the exponential priors are characterized by a
single parameter, the mean, they are far more easily constrained by
the observation(s).  But, for completeness and for illustrative
purposes we consider here two Gaussian distributions each with a mean of
unity, but having differing dispersions. We make no claim that these two
parameters can be constrained by a single observation. We simply want
to illustrate quantitatively that the uniform prior at least (and the
exponential prior in its extreme) both encompass the results for these
Gaussian priors as well. That is, by adopting the {\it g-factors} for
a uniform prior one will also encompass the confidence intervals
derived for a variety of Gaussian distributions have the same (unit)
mean and certainly any dispersion less than 0.5.

The results for these additional priors are given graphically in
Figures 3 and 4, and in Columns 5 and 6 of Table 1.

\section{Discussion and Conclusions}

Our interest in this formalism started with a desire to characterize
luminosity variations of astronomical objects (in the first instance
through their first two moments, chosen to be their means and
amplitudes) having ahighly restricted number of observations. Here we
have taken that thought experiment to its limit of a singular
observation. For an underlying distribution of finite extent (or
duration) the dispersion and the full amplitude are equivalent,
differing only by a constant scale factor in any given
case. Accordingly, our derivation of the variance of a population
based on a single observation is conceptually equivalent to Gott's
(1993) derivation of future longevity (that is, Gott's longevity is a
one-directional semi-amplitude) for an object based on a ``single''
observation of its present age.

When dealing with physical observations certain assumptions are
tacitly taken for granted. The obvious assumptions are, that those 
quantities are non-negative, and that the underlying population from
which they are drawn does not have an infinite range (in time, mass,
energy, size, {\it etc.})  However, those same assumptions carry
additionally useful (prior) quantitative information that can be used
to constrain limits on observations as they are obtained. This paper
had the intent of making those assumptions explicit and then
formalizing the mathematical consequences expressed as confidence
intervals on the underlying population as derived from one observation

While it was our hope that this intentionally short contribution might
stimulate others into finding applications not obvious to the author,
the referee requested that examples be given of how this formalism
might be applied to astronomy. In keeping with the spirit of this
paper we offer a modest example from the past, and predict that there
will be future examples; all of this based on a sample of N = 1.

Laplace developed his ``rule of succession'' when confronted with a
question as to the mathematical (not the physical) probability that
the Sun will rise tomorrow given its past (statistical)
performance. After observing N events, Laplace derived that the
probability of the next occurrence was $(N+1)/(N+2)$. This would
suggest that on the first day (N = 0) the probability of the Sun rising
was actually 50:50. On the second day (N = 1) the probability would
have gone up to 66\%, and so on. Gott (1993) asked a similar question,
not just about the next occurrence of something that has a past
persistence, but about the sum of all future occurrences. How long will
a thing last, given that we know how old it is now? Gott was
reformulating Laplace's question to be, if we have a single measurement
(N = 1) of the age of something, what can one say about its total
lifetime (its full amplitude), or rather its future longevity (a
semi-amplitude). Of course, any such prediction is best described in
terms of probabilities, and so rather than predicting a firm lifetime
based on a precise age, Gott predicted a forward-looking probability
distribution (expressed as a variance) based on a backward-looking age.

One could argue that Gott actually required two observations: one of
the time at which something began and another of the time at which the
prediction was being made. This may be seen as quibbling but it is, in
fact, equivalent to our physical prior, state above, that none of the
quantities to which this method applies can drop below zero or
become infinite in amplitude (mass, luminosity, time, etc).
 
Grounded with post-dictions on the longevity of the Soviet Union and
the Berlin Wall, Gott went on to predict the longevity of a variety of
things astronomically big and small: from the expected demise of {\it
Nature} magazine itself (somewhere between 3.15 and 4,800 years), to
the probability that we will end as a civilization (in 5,100 to 8
million years with 95\% probability), or colonize the Galaxy (the odds
are against it). Each of these predictions were based on a single
observation, N = 1 and a uniform prior.

It is quite clear that the uniform, non-negative prior distribution of
data points discussed above is at one extreme (of simplicity, or of
ignorance.)  However, this extreme is also rather inclusive. If the
true range of the underlying distribution is smaller, more centrally
peaked, or more skewed toward the upper bound of the distribution
function, than a uniform distribution, then their confidence intervals
will also be smaller than those calculated for a uniform prior; under
those conditions the uniform prior is likely to provide a conservative
{\it upper} bound on the uncertainty.

\subsection{Is This Just Another Name for Modeling?}

Finally, we suggest that aspects of the scientific enterprise as a
whole, as practiced by many astronomers in interpreting observational
data, might simply be a generalization of the the Sigma One
methodology discussed here. Seen in that retrospective light, Sigma
One becomes a fairly benign and low-level form of what would otherwise
be called ``modeling''.

Consider the observation of a color-magnitude diagram for a composite
stellar population, in a nearby galaxy, say. Based on that single
observation one could ask what the magnitude and color of any given
star might be on the next exposure (whenever that may be.) Depending
on the amount of prior knowledge about the underlying distribution
function for that star one could make a prediction. Indeed we do this
all the time. It is known that intrinsic variables (Cepheids, Miras,
RR Lyrae stars, etc.)  occupy fairly well-delineated regions of the
color-magnitude diagram. Armed with known amplitudes and timescales
one could invoke those distribution functions with their specific
means and variances to predict the expected variance in those selected
stars. Stars in regions not known to be variable on those same
time-scales would have different priors used to predict their means
and variances.

But all of this could also be recast into a very different form of the
underlying distribution function, in the case where extremely long
(astronomically long) timescales are being considered.  The predictive prior
for a single observation would then become stellar evolution theory
itself. That is, given a star observed (once) today at a given place
in the color-magnitude diagram, what is its color and magnitude
distribution function integrated over its projected future existence?
And then how might the ensemble change with time?  We apparently have
no problem in undertaking population synthesis modeling, for example,
taking a single integrated spectrum and/or a single color-magnitude
diagram and extrapolating it to encompass the entire life history
(backward and forward in time) of a given star and/or its associated
contemporary population (an entire galaxy).  So our point here is that
if we are comfortable extracting very complex ``moments'' (in time and
composition, etc.)  from single (but admittedly very rich)
observations by invoking very complicated priors (i.e., models), then
it should come as no surprise that it is possible to extract more than
just one moment (i.e., a mean and a variance at least) from a single
data point by assuming very simple priors (i.e., models), in the form
of well known, commonly invoked, but simple, distribution functions.

\acknowledgements

I thank Wendy Freedman for pointing out both the {\it prior} nature
and the {\it a posteriori} relevance of the Gott (1993) Nature paper.
study. Discussions with David Hogg were both stimulating and
illuminating, as always.  And finally, I would like to thank both the
Editor, Jay Gallagher for his patience in dealing with this paper as
it slowly evolved, and the anonymous referee who carefully read the
paper gave many helpful suggestions.

\noindent
\centerline{\bf References \rm}
\vskip 0.1cm

Gott, J.R., III, 1993, Nature, 363, 315

Jaynes, E.T. 2003 ``Probability Theory: The Logic of Science''
ed. G.L. Bretthorst, Cambridge University Press, Cambridge

Madore, B.F. \& Freedman, W.L. 2005, ApJ., 630, 1054

\vfill\eject
\vskip 0.75cm


\clearpage 
\begin{figure}
\centering
\includegraphics [width=12cm, angle=270] {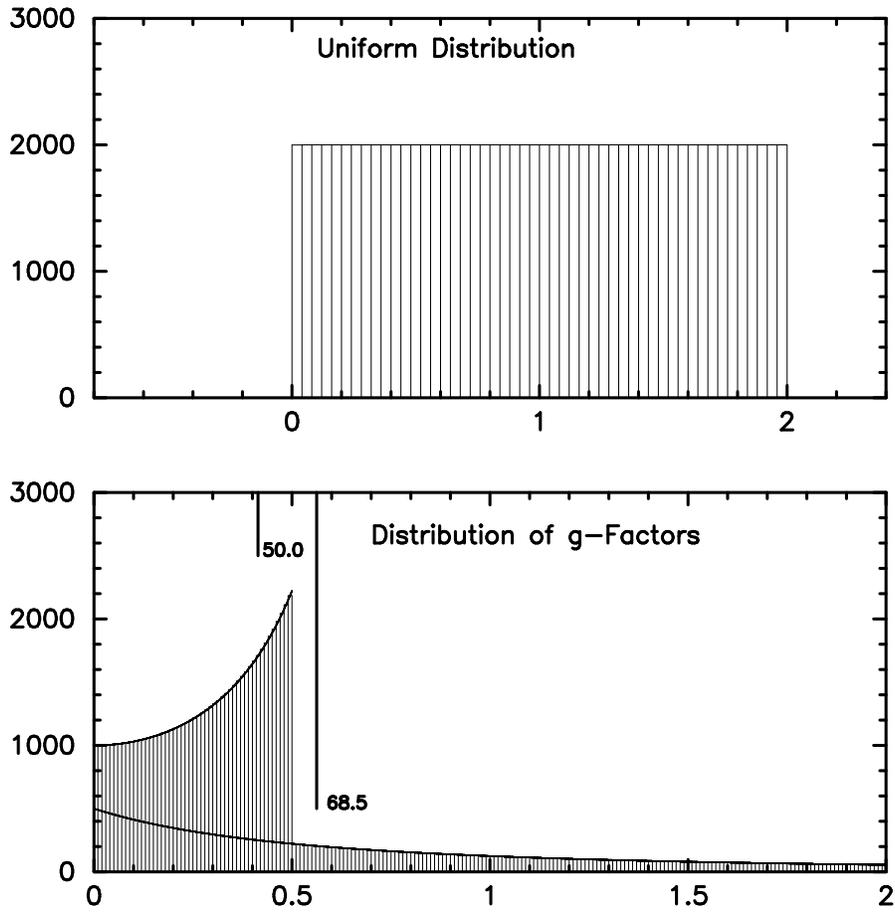}
\caption{ -- Uniform Prior. The upper panel shows the input
distribution of points having a uniform distribution with a mean of 1
and a width of 2. The lower panel shows the cumulative distribution of
the multiplicative {\it g-factors} (as a function of the {\it
g-factor} itself) needed to convert random samples taken from the
upper panels into the mean. The {\it g-factors} required to give 50\%
and 68.5\% (one sigma) confidence intervals for this underlying
(uniformly distributed) sample are given as vertical bars and labeled
accordingly }
\end{figure}

\clearpage 
\begin{figure}
\centering
\includegraphics [width=12cm, angle=270] {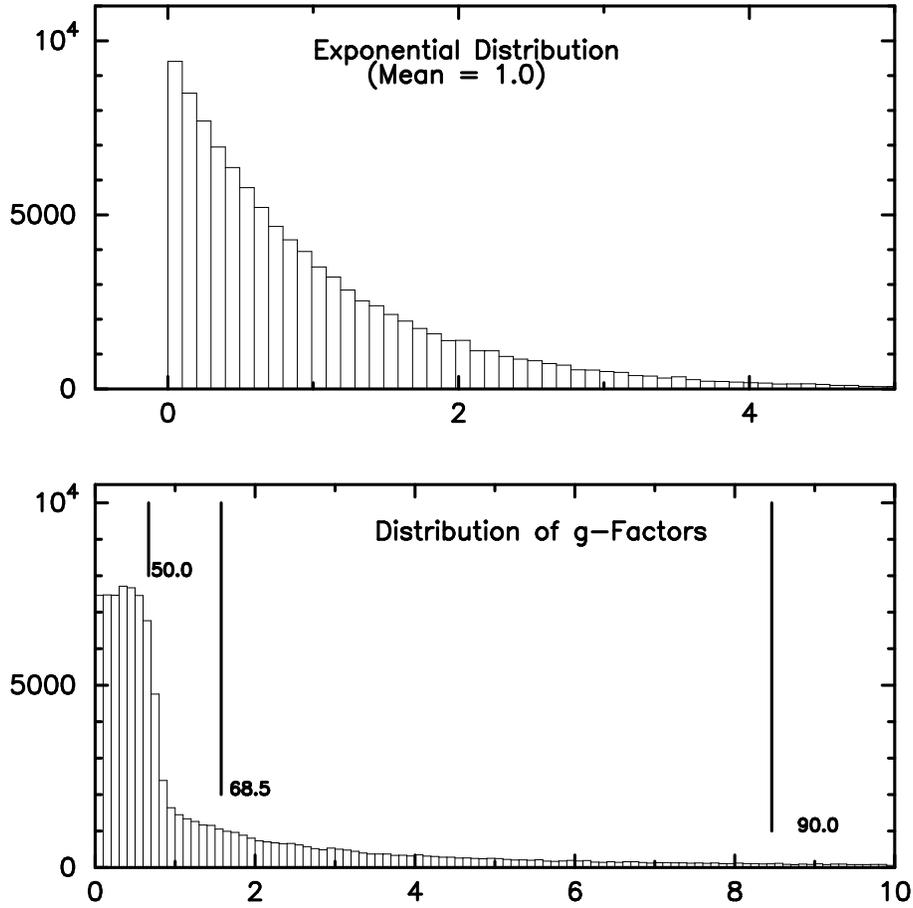}
\caption{ -- Exponential Prior. The upper panel shows the input
distribution of points having an exponential distribution with a mean
of 1. The lower panel shows the cumulative distribution of the
multiplicative {\it g-factors} (as a function of the {\it g-factor}
itself) needed to convert random samples taken from the upper panels
into the mean. The {\it g-factors} required to give 50\%, 68.5\% (one
sigma) and 90.0\% confidence intervals for this underlying
(exponentially distributed) sample are given as vertical bars and
labeled accordingly }
\end{figure}

\clearpage 
\begin{figure}
\centering
\includegraphics [width=12cm, angle=270] {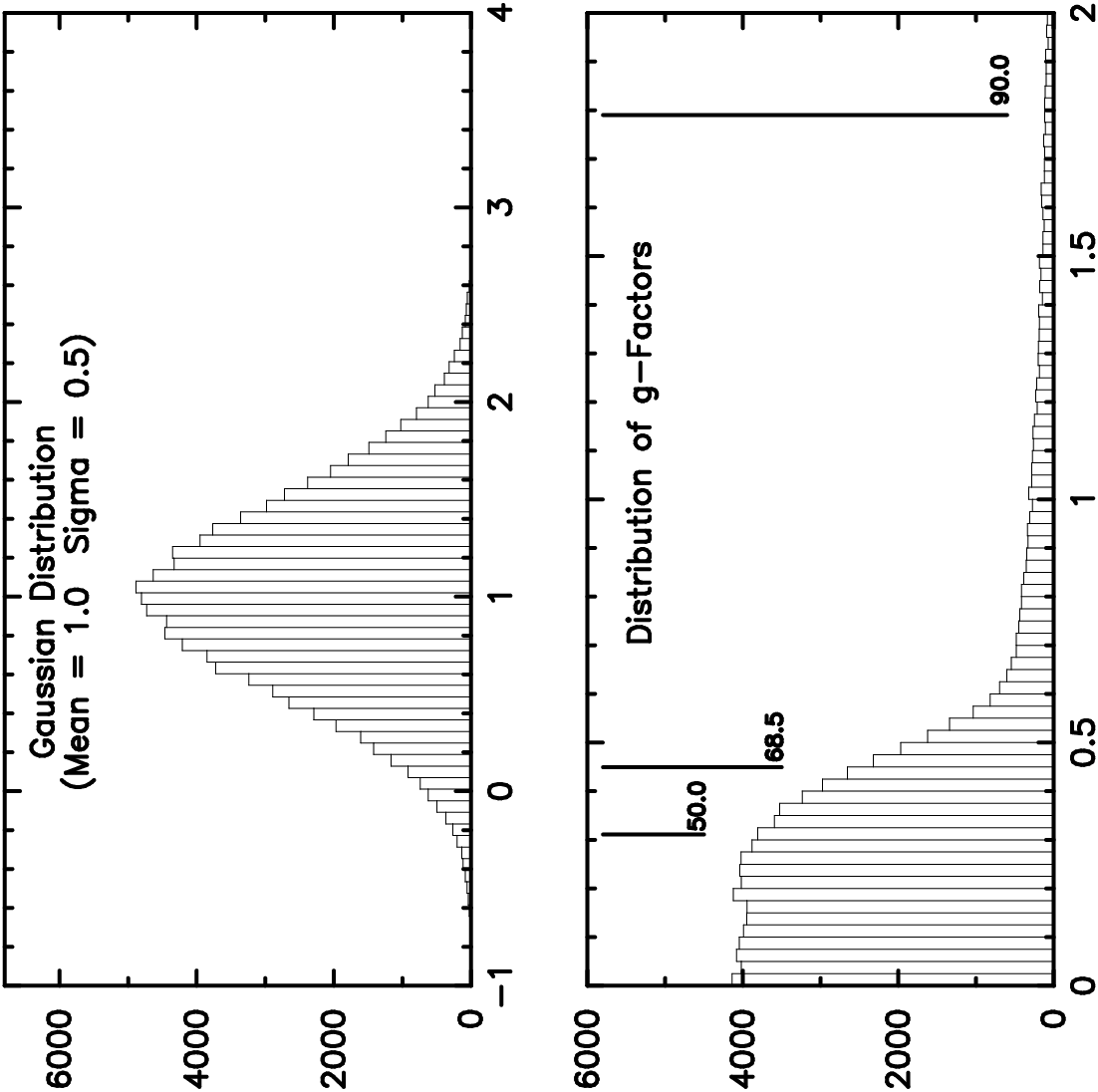}
\caption{ -- Gaussian ($\sigma$ = 0.5) Prior. The upper panel shows
the input distribution of points having a Gaussian distribution with a
mean of 1 and a sigma of 0.5. The lower panel shows the cumulative
distribution of the multiplicative {\it g-factors} (as a function of
the {\it g-factor} itself) needed to convert random samples taken from
the upper panels into the mean. The {\it g-factors} required to give
50\%, 68.5\% (one sigma) and 90.0\% (two sigma) confidence intervals
for this underlying (normally distributed) sample are given as
vertical bars and labeled accordingly }
\end{figure}

\clearpage 
\begin{figure}
\centering
\includegraphics [width=12cm, angle=270] {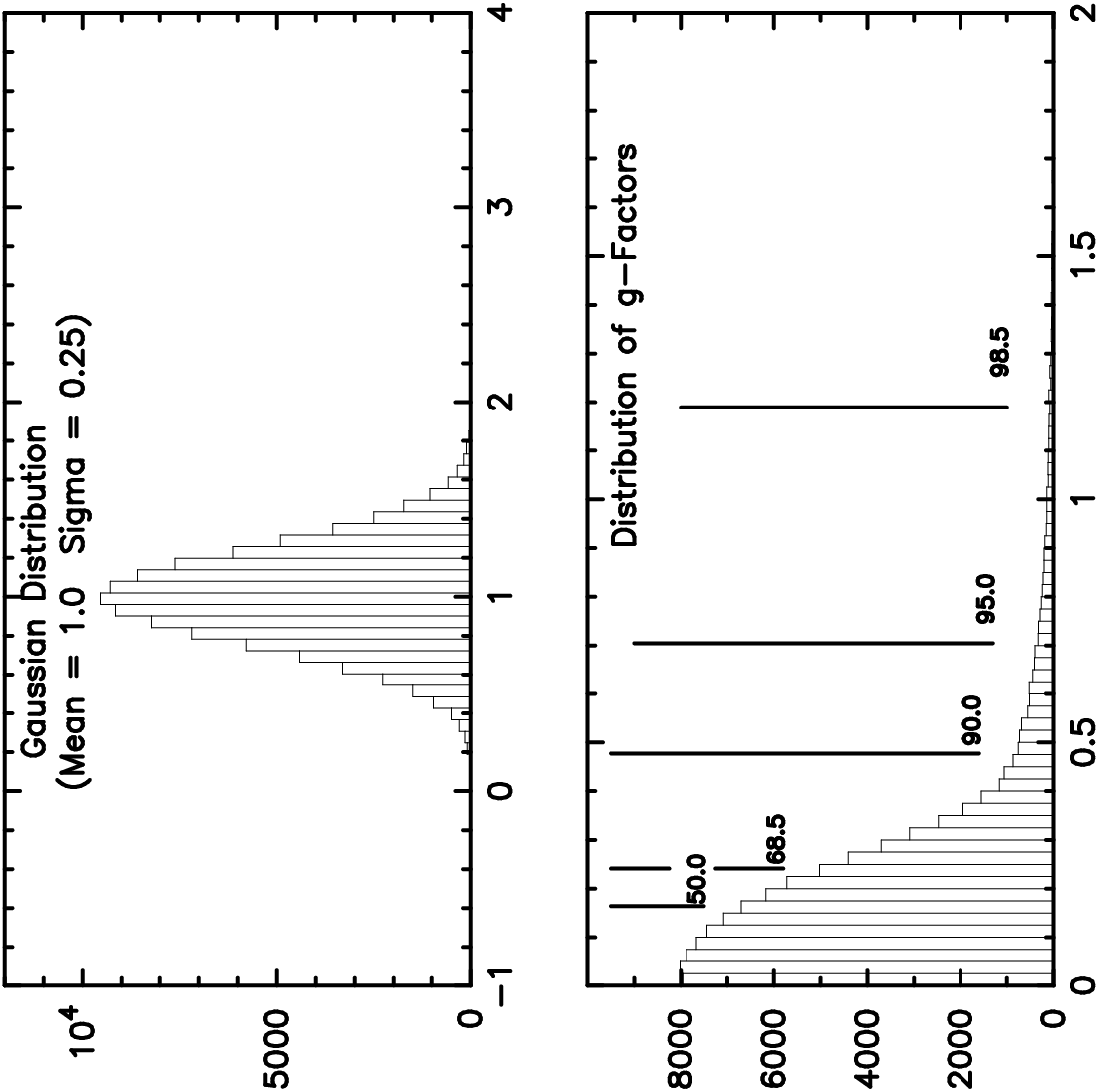}
\caption{ -- Gaussian ($\sigma$ = 0.25) Prior. The upper panel shows
the input distribution of points having a Gaussian distribution with a
mean of 1 and a sigma of 0.25. The lower panel shows the cumulative
distribution of the multiplicative {\it g-factors} (as a function of
the {\it g-factor} itself) needed to convert random samples taken from
the upper panels into the mean. The {\it g-factors} required to give
50\%, 68.5\% (one sigma), 90.0\% (two sigma) and 98.5\% (three sigma)
confidence intervals for this underlying (normally distributed) sample
are given as vertical bars and labelled accordingly }

\end{figure}

\vfill\eject
\noindent

\clearpage

\begin{deluxetable}{ccccc}
\tablecaption{Confidence Intervals and g-Factors for Selected
Priors\label{tbl-1}}
\tablewidth{5.5truein}
\tablehead{
\colhead{Confidence} & \colhead{Uniform} & \colhead{Exponential} & \colhead{Gaussian(0.50)} & \colhead{Gaussian(0.25)} \cr
\colhead{Interval} & \colhead{g-Factor}  & \colhead{g-Factor} & \colhead{g-Factor} & \colhead{g-Factor}}
\startdata
50.0\% & ~$\pm$~0.414 & ~$\pm$~0.67 & ~$\pm$~0.31 & ~$\pm$~0.16 \\
68.5\% & ~$\pm$~0.563 & ~$\pm$~1.60 & ~$\pm$~0.45 & ~$\pm$~0.24 \\
90.0\% & ~$\pm$~4.000 & ~$\pm$~8.5~ & ~$\pm$~1.78 & ~$\pm$~0.48 \\
95.0\% & ~$\pm$~9.000 & $\pm$~19.~~ & ~$\pm$~4.0~ & $\pm$~0.70 \\
98.5\% & $\pm$~32.333 & $\pm$~65.~~ & $\pm$~14.~~ & $\pm$~1.2~ \\
\enddata
\end{deluxetable}

\end{document}